

\font\mybb=msbm10 at 12pt
\def\bb#1{\hbox{\mybb#1}}
\def\Z {\bb{Z}}
\def\R {\bb{R}}

\tolerance=10000
\input phyzzx

 \def\unit{\hbox to 3.3pt{\hskip1.3pt \vrule height 7pt width .4pt \hskip.7pt
\vrule height 7.85pt width .4pt \kern-2.4pt
\hrulefill \kern-3pt
\raise 4pt\hbox{\char'40}}}
\def\II{{\unit}}

\def\half{{\textstyle {1 \over 2}}}
\def\V{{\cal V}}
\def\ZZ {{\cal Z}}
\def\RR {{\cal R}}

\REF\FK{{\sl Kac-Moody and Virasoro algebras: A Reprint Volume for Physicists},
eds. P. Goddard and
D. Olive, World Scientific (1988).}
\REF\Narain {K.S.  Narain, Phys. Lett. {\bf B169} (1986) 41.}
\REF\covlat{W. Lerche, A.N. Schellekens and N.P. Warner,  Phys. Rep. {\bf 177}
(1989) 1.}
\REF\goddard{Bluhm, P. Goddard and L. Dolan, Nucl. Phys.
 {\bf B289}  (1987), 364; {\bf B309} (1988) 330. }
\REF\kap{L.J. Dixon, V. Kaplanovsky and C. Vafa, Nucl. Phys. {\bf B294}  (1987)
43; W. Lerche, B.E.N. Nilsson, A.N. Schellekens and N.P. Warner,   Nucl. Phys.
{\bf B294} (1987) 136.}
\REF\WO{E. Witten and D. Olive, Phys. Lett. {\bf B78} (1978) 97.}
\REF\GH {G.W. Gibbons and C.M. Hull, Phys. Lett. {\bf 109B} (1982) 190.}
\REF\Cer{A. Ceresole, R. D'Auria, S. Ferrara and A. Van Proeyen,
{\it Duality Transformations in Supersymmetric Yang-Mills Theories coupled
to Supergravity}, preprint  CERN-TH 7547/94, POLFIS-TH. 01/95, UCLA 94/TEP/45,
KUL-TF-95/4, hep-th/9502072.}
\REF\HT{C.M. Hull and P.K. Townsend, Nucl. Phys. {\bf B438} (1995) 109.}
\REF\Witten{E. Witten, Nucl. Phys. {\bf B444} (1995) 161.}
\REF\Stromb{A. Strominger, {\it Massless black holes and conifolds in string
theory}, hep-th/9504090.}
\REF\vafa{C. Vafa, {\it A stringy test of the fate of the conifold},
hep-th/9505023.}
\REF\GMS{B.R. Greene, D.R. Morrison and A. Strominger, {\it Black hole
condensation
and the unification of string vacua}, hep-th/9504145.}
\REF\KV{S. Kachru and C. Vafa, {\it Exact results for N=2 compactifications of
heterotic strings}, hep-th/9505105.}
\REF\SW{E. Witten and N. Seiberg, Nucl. Phys. {\bf B426} (1994) 19.}
\REF\DNP{M.J. Duff, B.E.W. Nilsson and C.N. Pope, Phys. Lett {\bf 129B}
(1983) 39; M.J. Duff and B.E.W. Nilsson, Phys. Lett. {\bf B175} (1986) 39.}
\REF\PKT{P.K. Townsend, Phys. Lett. {\bf B350}, (1995) 184.}
\REF\HS {G.T. Horowitz and A. Strominger, Nucl. Phys. {\bf B360} (1991)
197.}
\REF\PKTb{P.K. Townsend, Phys. Lett. {\bf 354B} (1995) 247.}
\REF\Gu {R. G{\" u}ven, Phys. Lett. {\bf 276B} (1992) 49.}
\REF\BST{E. Bergshoeff, E. Sezgin and P.K. Townsend, Phys. Lett. {\bf 189B}
(1987) 75; Ann. Phys. (N.Y.) {\bf 185} (1988) 330.}
\REF\sei{N. Seiberg,  Nucl.Phys. {\bf B431}  (1994) 551; {\bf  B435} (1995)
129; K. Intriligator and N. Seiberg,  hep-th/9503179.}
\REF\Sen {A. Sen, Nucl. Phys. {\bf B404} (1993) 109; Phys. Lett. {\bf 303B}
(1993); Int. J. Mod. Phys. {\bf A8} (1993) 5079; Mod. Phys. Lett. {\bf
A8} (1993) 2023;     Int. J. Mod. Phys. {\bf A9} (1994) 3707..}
\REF\ssen{A. Sen, Phys. Lett. {\bf B329} (1994) 217.}
\REF\HL {J. Harvey, J. Liu, Phys. Lett. {\bf B268} (1991) 40.}
\REF\FILQ{A. Font, L. Ibanez, D. Lust and F. Quevedo, Phys. Lett. {\bf B249}
(1990) 35; S.J. Rey, Phys. Rev. {\bf D43}  (1991) 526.}
\REF\GM{G.W. Gibbons and K. Maeda, Nucl. Phys. {\bf B298} (1988) 741.}
\REF\DGHR {A. Dabholkar, G.W. Gibbons, J.A. Harvey and F. Ruiz-Ruiz,
Nucl. Phys. {\bf B340} (1990) 33.}
\REF\CHS {C. Callan, J. Harvey and A. Strominger, Nucl. Phys. {\bf B359}
(1991) 611.}
\REF\DL {M.J. Duff and J.X. Lu, Nucl. Phys. {\bf B354} (1991) 141;
Phys. Lett. {\bf 273B} (1991) 409.}
\REF\DLb {M.J. Duff and J.X. Lu, Nucl. Phys. {\bf B416} (1993)  301.}
\REF\khuria{R.R. Khuri,
Phys. Lett. {\bf  B259} (1991) 261; Nucl. Phys. {\bf B387} (1992) 315.}
\REF\talk{C.M. Hull, talk given at Strings '95, USC, March 1995.}
\REF\GKLTT{G.W. Gibbons, D. Kastor, L. London, P.K. Townsend and J.
Traschen, Nucl. Phys. {\bf B416} (1994) 850.}
\REF\Kallosh {R. Kallosh, {\it Black hole multiplets and spontaneous breaking
of
local supersymmetry}, preprint hep-th/9503029.}
\REF\dR {M. de Roo, Nucl. Phys. {\bf B255} (1985) 515.}
\REF\berg{E. Bergshoeff, I.G. Koh and E. Sezgin, Phys. Lett. {\bf 155B} (1985)
71.}
\REF\SS {J.H. Schwarz and A. Sen, Nucl. Phys. {\bf B411} (1994) 35; Phys.
Lett. {\bf 312B} (1993) 105.}
\REF\AM{P.S. Aspinwall and D.R. Morrison, {\it U-Duality and Integral
Structures},
hep-th/9505025.}
\REF\bergort{E. Bergshoeff, C.M.Hull and T. Ortin, preprint UG--3/95,
QMW--PH--95--2, hep-th/9504081, Nucl. Phys. {\bf B} {\it in press}.}
\REF\ferras{ S. Cecotti, S. Ferrara and L. Girardello, Int. Journ. Mod. Phys.
{\bf A4} (1989) 2475; S. Ferrara and S. Sabharwal, Class. Quan. Grav. {\bf 6}
(1989) L77.}
\REF\duff{M.J. Duff, CTP-TAMU-49/94, hep-th/9501030.}
\REF\GETC {G.W. Gibbons, Nucl. Phys. {\bf B207} (1982) 337; G.W. Gibbons and
K. Maeda, Nucl. Phys. {\bf B298} (1988) 741; D. Garfinkle, G.T. Horowitz and
A. Strominger, Phys. Rev. {\bf D43}, (1991) 3140; C.F. Holzhey and F. Wilczek,
Nucl. Phys. {\bf B380} (1992) 447.}
\REF\Ort{R. Kallosh, A. Linde, T. Ortin, A. Peet, A.
Van Proeyen, Phys.Rev. {\bf D46} (992) 5278.}
\REF\Kalort{R. Kallosh and  T. Ortin, Phys.Rev. {\bf D48}  (1993) 742. }
\REF\GP {G.W. Gibbons and M.J. Perry, Nucl. Phys. {\bf B248} (1984) 629.}
\REF\Druff{ M.J. Duff and J. Rahmfeld, Phys. Lett. {\bf  B345} (1995) 441,
hep-th/9406105.}
\REF\SGP {R. Sorkin, Phys. Rev. Lett. {\bf 51} (1983) 87 ; D. Gross and
M. Perry, Nucl. Phys. {\bf B226} (1983) 29.}
\REF\GHL {J. Gauntlett, J. Harvey and J. Liu, Nucl. Phys. {\bf B409}
(1993) 363; J. Gauntlett and J. Harvey, {\it S-Duality and the spectrum of
magnetic monopoles in heterotic string theory}, preprint EFI-94-36.}
\REF\GHT {G.W. Gibbons, G.T. Horowitz and P.K. Townsend, Class. Quantum Grav.
{\bf 12} (1995) 297.}
\REF\DS {M.J. Duff and K.S. Stelle, Phys. Lett. {\bf 253B} (1991) 113.}
\REF\DGT {M.J. Duff, G.W. Gibbons and P.K. Townsend, Phys. Lett. {\bf 332 B}
(1994) 321.}
\REF\Oopen{C.M.Hull, {\it String-String Duality in Ten Dimensions},
hep-th/9506194.}
\REF\BKS{E. Bergshoeff, I.G. Koh and E. Sezgin, Phys. Rev. {\bf D32} (1985)
1353.}
\REF\AGIT{J. A. de Azc{\' a}rraga, J.P. Gauntlett, J.M. Izquierdo and
P.K. Townsend, Phys. Rev. Lett. {\bf 63} (1989) 2443.}


\Pubnum{ \vbox{ \hbox {QMW-95-10} \hbox{R/95/11}  \hbox{hep-th/9505073}} }
\pubtype{}
\date{May 1995, revised August 1995}

\titlepage

\title {\bf ENHANCED GAUGE SYMMETRIES IN SUPERSTRING THEORIES}

\author{C.M. Hull}
\address{Physics Department,
Queen Mary and Westfield College,
\break
Mile End Road, London E1 4NS, U.K.}
\andauthor{P. K. Townsend}
\address{DAMTP, University of Cambridge,
\break
Silver Street, Cambridge CB3 9EW,  U.K.}

\eject

\abstract{Certain four-dimensional $N=4$ supersymmetric theories have special
vacua
in which massive charged vector supermultiplets become massless, resulting in
an
enhanced non-abelian gauge symmetry. We show here that any two $N=4$ theories
having
the same Bogomolnyi spectrum at corresponding points of their moduli spaces
have the same
enhanced symmetry groups.
In particular, the $K_3\times T^2$ compactified type II string is argued to
have the same
enhanced symmetry groups as the $T^6$-compactified heterotic string, giving
further
evidence
for our conjecture that these two  string theories are equivalent.  A feature
of the enhanced symmetry
phase is that for every electrically charged state whose mass tends to zero as
an enhanced symmetry
point is approached,  there are magnetically charged  and dyonic states whose
masses also tend to
zero, a result that applies equally to N=4 super Yang-Mills theory. These extra
non-perturbative massless states in the
$K_3$ compactification result from $p$-branes wrapping around collapsed
homology two-cycles of $K_3$.
Finally, we
show how membrane `wrapping modes' lead to symmetry
enhancement in D=11 supergravity,
providing further evidence that the $K_3$-compactified D=11
supergravity is the effective field theory of the strong coupling limit of the
$T^3$-compactified heterotic string.}

\endpage
\pagenumber=1


\chapter{Introduction}

Non-abelian gauge symmetry is the crucial ingredient of the standard model of
particle physics. It is hoped that the gauge symmetries of the standard model
will eventually be understood in terms of some more fundamental theory that
unifies the electro-weak and strong forces with gravity, such as superstring
theory. Various ways in which non-abelian gauge symmetry can emerge from such
unified theories have been proposed in the past, many of them involving extra
compact dimensions. The most obvious way that this can happen is if the
non-abelian gauge symmetry is already an ingredient of the candidate
fundamental
theory. An example is the $E_8\times E_6$ gauge symmetry of the heterotic
string
theory compactified on a Calabi-Yau manifold, which has its origin in the
$E_8\times E_8$ gauge symmetry of the ten-dimensional theory. Another
possibility is the Kaluza-Klein (KK) mechanism in which the gauge symmetry
arises from the isometry group of an internal compact space. A third way is
the Halpern-Frenkel-Ka\v c (HFK) mechanism for symmetry enhancement in
toroidally-compactified string theories at weak string coupling (see, for
example, [\FK]). The simplest example of the HFK mechanism occurs for the
bosonic string compactified on a circle of radius $R$; for generic values of
$R$, there is only a $U(1)$ gauge symmetry of Kaluza-Klein origin but at the
self-dual radius this is enhanced to $SU(2)$. Another example is the toroidal
compactification of the heterotic string to four dimensions for which the gauge
group is $U(1)^6\times G$. The group $G$ is $U(1)^{22}$ for generic
compactifications, but is a non-abelian group of rank $22$ [\Narain] at special
points in the moduli space of the compactification. A form of the HFK mechanism
is also applicable to the type II string in the covariant lattice approach
[\covlat], in which the fermionic constructions  of [\goddard] are recovered as
special cases. However, in the type II string this mechanism does not lead to
realistic theories [\kap].

In these string theory examples, the HFK symmetry enhancement is due to massive
spin one states in the spectrum becoming massless at special points of the
moduli space. Some of these massive states can be interpreted as due to string
winding modes, and the possibility of symmetry enhancement by this mechanism
has therefore been seen as an intrinisically `stringy' one. However, we will
show that symmetry enhancement in fact occurs for any theory which has an $N=4$
supersymmetric low-energy effective action and which has the requisite massive
states; these might arise either as perturbative states (e.g. Kaluza-Klein
modes) or as solitons. The symmetry enhancement is understood in terms of the
effective field theory and is independent of the nature of the compactifying
space, e.g. whether or not it has isometries.

Mass eigenstates of any four-dimensional theory with $N\ge 2$ supersymmetry can
be chosen to be simultaneous eigenstates of the central charges appearing in
the
supersymmetry algebra, and the mass of any central charge eigenstate is bounded
by the values of  its central charges [\WO]; this is generally called the
`Bogomolnyi bound'. In the supergravity context, i.e. for theories with local
supersymmetry, the $N(N-1)$ central charges are the electric and magnetic
charges associated with the $N(N-1)/2$ abelian gauge fields in the graviton
supermultiplet. In this case, the Bogomolnyi bound  [\GH] is a lower bound on
the ADM energy of an asymptotically-flat spacetime given in terms of the total
electric and magnetic charges, and the asymptotic values of the scalar fields.
These scalar field values parameterise the possible classical vacua, and can
be
interpreted as coupling constants of the theory. Classically, the electric and
magnetic charges can be chosen arbitrarily but in the quantum theory the
Dirac-Schwinger-Zwanziger (DSZ) quantization condition restricts their values
to a lattice, given the existence of electrically and magnetically charged
states of each type. The mass of any `Bogomolnyi state' saturating the
Bogomolnyi bound is then determined by its location on the charge lattice and
the choice of vacuum, i.e. the scalar field expectation values.

For pure $N\ge2$ supergravity theories without matter coupling, the only zero
mass states saturating the bound are those with zero electric and magnetic
charges. A new feature emerges for $N\ge2$ Maxwell/Einstein supergravity
theories, i.e. $N\ge2$ supergravity coupled to abelian vector supermultiplets,
as a result of the additional electric and magnetic charges. As shown in an
explicit $N=2$ example [\Cer], there can be charged  states that saturate the
bound but whose masses vanish at special points of the scalar field target
space, i.e. for special choices of the vacuum. Some of these massless states
have spin one and this results in an enhancement of the abelian symmetry to a
non-abelian group. However, these $N=2$ results only hold in a semi-classical
approximation and the masses will in general receive quantum corrections
[\Cer],
even when the $N=2$ theory is the low-energy effective action for a superstring
theory, such as the type II string compactified on a Calabi-Yau manifold or the
heterotic string compactified on $K_3 \times T^2$. Here we shall concentrate on
theories with at least $N=4$ supersymmetry as in such cases the masses and
charges of Bogomolnyi states are believed to receive no quantum corrections,
thus justifying semi-classical reasoning based on a non-renormalizable
effective theory. We shall show that for any theory for which the effective
four-dimensional field theory is an $N=4$ Maxwell/Einstein supergravity, a
non-abelian gauge symmetry will appear at special points of the sigma-model
target space, provided only that there exist Bogomolnyi states in the spectrum
carrying the appropriate charges.

For the toroidally compactified heterotic string the effective four dimensional
field theory is, for generic compactifications,  N=4 supergravity coupled to 22
abelian vector multiplets. The required electrically charged Bogomolnyi states
arise in string  perturbation theory, so that there exist points in moduli
space of enhanced gauge symmetry, in agreement with the results of the HFK
mechanism. More significantly, the same supergravity theory is also the
effective four dimensional field theory of either the type IIA or the type IIB
superstring compactified on $K_3\times T^2$. This fact had been noted on
various occasions in the past but was considered to be merely a coincidence. We
argued in [\HT] that the massive spectrum could also be the same once account
was taken of non-perturbative states resulting from wrapping $p$-brane solitons
of the ten-dimensional string round homology cycles of $K_3\times T^2$, and a
study of these `wrapping modes' led us to conjecture that the toroidally
compactified heterotic and $K_3\times T^2$ compactified type II superstrings
are actually equivalent. We pointed out that this conjecture implies the
existence of special points of enhanced symmetry in the moduli space of
$K_3\times T^2$ compactifications of type II superstrings and we suggested that
this might occur as a result of some non-perturbative mechanism involving
$p$-brane solitons.

This suggestion, and the conjecture, became much more plausible after Witten
pointed out that for every point in the heterotic string moduli space at which
the symmetry was enhanced, the corresponding point in the $K_3$ moduli space
was associated with a limit in which certain homology two-cycles of $K_3$
[\Witten] collapse to zero area. To see the relevance of this observation, we
recall the identification in [\HT] of massive modes carrying 22 of the 28
electric charges with the wrapping modes of the 2-brane soliton of the type
IIA superstring around the 22 homology cycles of $K_3$. Since the mass of these
states is proportional to the area of the two-cycle the mass should vanish when
the two-cycle collapses to zero area. Moreover, since these states are BPS
saturated they belong to ultrashort multiplets which can turn into massless
vector supermultiplets as the mass goes to zero, as required for symmetry
enhancement.

One potential problem with this mechanism of symmetry enhancement, which was
discussed  in the context of N=2 superstring compactifications in [\Stromb],
is that the concept of a $p$-brane soliton is intrinsically semi-classical; the
description of states as $p$-brane wrapping modes makes sense only
when the $p$-brane tension is large and the $p$-cycles around which they are
wrapped are large relative to the scale set by the $p$-brane core. It is clear
that one needs additional input from supersymmetry to justify the extension of
$p$-brane wrapping modes to the limit in which the $p$-cycle collapses. In the
case of N=2 superstring compactifications, the effective supergravity theory is
not uniquely determined by supersymmetry so that considerations of
renormalization must play a role. The picture now emerging [\vafa,\GMS,\KV] is
of a stringy generalization of the work of Seiberg and Witten [\SW] and
Ceresole
{\sl et al.} [\Cer]. Here we shall be examining this mechanism for symmetry
enhancement in the context of N=4 superstring compactifications. In this case,
the effective supergravity theory {\it is} determined uniquely by supersymmetry
(once the spectrum of massless states and the Yang-Mills gauge group are
known).

As we shall see, for theories with $N=4$ supersymmetry the masses of
Bogomolnyi states are also determined entirely by the low-energy effective
field
theory. Thus, once the existence of a massive Bogomolnyi state has been
established at some particular point in the moduli space of vacua, {\it by
whatever means}, its mass at other points in the moduli space is determined by
the effective low-energy supergravity theory. In particular, the mass of
certain  Bogomolnyi states must vanish at special points in this moduli space
purely as a consequence of N=4 supersymmetry. Since these states must fill out
vector supermultiplets the proof of symmetry enhancement in N=4 supersymmetric
theories rests on the existence of the relevant massive states at generic
points in moduli space. Although we provide a characterisation of the relevant
states in various superstring compactifications, we rely on semi-classical
methods to establish their existence in the spectrum when these states are not
in the perturbative string spectrum. As in our previous work [\HT] on the
non-perturbative Bogomolnyi spectrum, which we amplify here, we identify the
required states as $p$-brane wrapping modes

The results so obtained provide further evidence for the conjectured
equivalence of the $K_3\times T^2$ compactified type II string to the $T^6$
compactified heterotic string. However, it is possible that these two theories
have identical Bogomolnyi spectra but differ in their non-Bogomolnyi spectra.
Indeed, since supersymmetry provides no clues to the spectrum of states in
full, unshortened, supermultiplets it might seem unlikely that any two $N=4$
theories would be the same even if their spectra of shortened supermultiplets
were to coincide in every detail. Of course, it may be that there are
specifically `stringy' reasons, not directly related to supersymmetry, why the
non-Bogomolnyi spectra must also coincide, but it is also possible that there
is a sense in which {\it there are no non-Bogomolnyi states} in the full
theory,
despite their occurrence in perturbation theory. States occurring in
perturbation theory that do {\it not} saturate a Bogomolnyi bound (strong or
weak) can be expected to correspond to particles that become unstable either at
higher orders of perturbation theory or non-perturbatively because there would
appear to be nothing to prevent their decay into Bogomolnyi states. In this
case
they could not appear in an exact S-matrix.

Another theory for which the effective four-dimensional field theory is an N=4
supergravity is D=11 supergravity compactified on $K_3\times T^3$ [\DNP]. In
fact, this effective field theory is identical to that of the $K_3\times T^2$
compactified type II string. Until recently this was thought to be just another
coincidence, but in [\HT] it was pointed out that, {\it when account is taken
of the $p$-brane solitons in the higher dimension}, the soliton spectrum in
four
dimensions is the same for D=11 supergravity compactified on $B\times T^3$ as
for a type II superstring compactified on $B\times T^2$, where $B$ is either
$T^4$ or $K_3$. This result hinted at an 11-dimensional interpretation of the
D=10 type IIA superstring theory but it was not clear then how such an
interpretation could be consistent with the fact that the critical dimension of
perturbative string theory is D=10. It was pointed out in [\PKT] that this
difficulty is resolved non-perturbatively by the existence in type IIA string
theory of the BPS saturated extreme black holes, found earlier by Horowitz and
Strominger [\HS], since the corresponding quantum states can be interpreted as
towers of massive Kaluza-Klein states. It was subsequently argued
in [\Witten], for similar reasons, that D=11 supergravity theory is the
effective field theory for the strong coupling limit of the type IIA
superstring. Specifically, the D=11 interpretation of the type IIA superstring
implies an identification of the string coupling constant with $R^{2/3}$ where
$R$ is the radius of the extra dimension. Thus, string  perturbation theory is
a
perturbation expansion about $R=0$, which explains why perturbative superstring
theory seems to require a spacetime of dimension D=10.

It was further argued in [\Witten] that the seven-dimensional field theory
obtained by $K_3$ compactification of D=11 supergravity is the effective action
for the strong coupling limit of the $T^3$ compactified heterotic
string. Further evidence for this proposition is the fact that the $T^3$
compactified heterotic string can be viewed [\PKTb] as a double-dimensional
reduction on $K_3$ of the D=11 fivebrane soliton of [\Gu]. If this proposition
is indeed true then it follows that the $K_3\times T^3$ compactified D=11
supergravity is the effective four-dimensional field theory for a limit of the
$T^6$ compactified heterotic string  in which one of the coupling constants
(moduli) becomes large. An obvious question that arises in this case is whether
the symmetry enhancement known to occur for the heterotic string can be seen
from the standpoint of D=11 supergravity. Since the latter is merely an
{\it effective} field theory it is not obvious that this should be possible.
Indeed, it is not possible within the confines of Kaluza-Klein theory but we
shall show that symmetry enhancement can be understood once $p$-brane solitons
of D=11 supergravity are included. If there were some  consistent quantum
theory
in 11-dimensions underlying D=11 supergravity one might consider this result to
be evidence of its equivalence, after compactification, to a superstring
theory.
There are reasons to believe [\PKT,\PKTb] that this 11-dimensional theory is
the D=11 supermembrane of [\BST], although we are a long way from understanding
its quantization.

A caution is in order before we continue with the derivation of the results
described above. In the toroidally compactified {\it bosonic} string the HFK
mechanism leads to (perturbative) symmetry enhancement in the effective
four-dimensional field theory but, since the effective non-abelian gauge theory
is asymptotically free, the conventional wisdom is that all particles carrying
non-abelian charge (which includes the massless non-abelian gauge bosons) are
confined. Thus, the extra massless states found in perturbation theory via the
HFK mechanism are not likely to be present in the full theory. If it were
possible to bypass perturbation theory and deal directly with the full quantum
string theory\foot{Because of the tachyon it is not clear that this exists.},
one might expect to find a transition to a confining phase at special points of
the moduli space, rather than the occurence of additional massless states at
these points.

The status of the HFK mechanism in the toroidally compactified heterotic string
is quite different because the effective non-abelian gauge theory at special
points of moduli space is not asymptotically free; in fact, the beta function
vanishes. Since confinement no longer operates to remove massless particles
with
non-abelian charge from the spectrum it might be thought that here, in contrast
to the bosonic string, the extra massless particles found in perturbation
theory
(either by the HFK mechanism or, as explained here, as a consequence of $N=4$
supersymmetry) indicate the existence of extra massless states in the full
theory. However, the infrared divergences due to unconfined non-abelian gauge
fields do not allow a standard interpretation of the Hilbert space in terms of
particles with definite charge quantum numbers. Instead, the existence of
vector fields whose masses tend to zero as one approaches special points in
moduli space signals a transition to a {\it non-abelian Coulomb phase} at these
points in which there is a non-abelian gauge symmetry associated with
long-range
forces; other theories with such phases have been investigated in [\sei].

As long as we are away from special points in moduli space there is no problem
in providing a standard particle interpretation for the spectrum. At first
sight
there would appear to be no difficulty in extrapolating this spectrum to those
special points at which some massive Bogomolnyi states become massless.
Symmetry
enhancement in the toroidally compactified heterotic string has so far only
been
analysed in string perturbation theory, with the result that electrically
charged Bogomolnyi perturbative string states become massless at special points
in moduli space. However, the heterotic string has non-perturbative
magnetically charged Bogomolnyi states arising from BPS monopole and dyonic
solutions of the low-energy theory [\Sen,\ssen,\HL]. We shall show that for
every electrically charged state whose mass tends to zero as one aproaches a
special point in moduli space, there is also a magnetically charged state and
(assuming S-duality [\FILQ]) an infinite set of dyonic ones whose masses also
approach zero. The interpretation of this is not clear, but possibly signals a
non-abelian Coulomb phase of a type rather different from those discussed in
[\sei]. However, it does show how symmetry enhancement might be consistent with
the conjectured S-duality of the heterotic string: whenever the mass
of a perturbative string state tends to zero, so do the masses of its
magnetically charged $SL(2,\Z)$ partners. Thus, even for the toroidally
compactified heterotic string we learn something more about the symmetry
enhancement mechanism from the analysis based on $N=4$ supersymmetry than we
learn from the HFK mechanism. A similar picture emerges for the type II string
on $K_3\times T^2$. In fact, since this feature of the enhanced symmetry phase
depends only on $N=4$ supersymmetry it applies equally to $N=4$ super
Yang-Mills
theory, i.e. as the Higgs expectation value tends to zero, {\it all} charged
massive states, including the magnetic monopoles and dyons, become massless
together. Whatever the correct interpretation of this may be, we wish to stress
here that it is a general feature. In particular, if, as argued in [\HT] and
again here, the Bogomolnyi spectrum of the $K_3\times T^2$ compactified type II
string is the same as that of the toroidally compactified heterotic string at
some generic points of their respective moduli spaces, then whatever happens to
the heterotic string at special points also happens to the type II superstring.

The thrust of this paper is that one can establish symmetry enhancement
in superstring compactifications to four dimensions preserving $N=4$
supersymmetry merely by an analysis of the low-energy effective field theory.
We
do this in two steps. First, as shown in the following section, the mass of a
Bogomolnyi state with a given charge vector is determined entirely by this
effective field theory, so that symmetry enhancement is the consequence of the
mere existence in the spectrum of certain states. We examine some consequences
of this result in the context of specific N=4 superstring compactifications in
section 3; in particular we extend the results of the perturbative HFK
mechanism
for the toroidally compactified heterotic string to the full
{\it non-perturbative} string theory. Second, as we explain in section 4, the
existence of the states relevant to symmetry enhancement for
$K_3$ compactifications can be deduced by consideration of the $p$-brane
soliton
solutions of the effective supergravity theory [\GM-\khuria] in a limit in
which semi-classical methods are reliable, because $N=4$ supersymmetry tells us
what happens to these states in all other regions of parameter space. In
section 5 we explain how these ideas can be used in the context of
compactifications of D=11 supergravity.

The results of sections 2 and 3 were announced at the Strings '95 conference
[\talk]. At the same meeting, the results of [\Witten] were also announced,
including some discussion of symmetry enhancement in compactified type II
theories.


\chapter{Symmetry Enhancement in $N=4$ Supergravity}

We now wish to see what can be learned directly from an analysis of the
effective four-dimensional theory for {\it any} theory which  has at least
$N=4$ local supersymmetry in $d=4$.  We shall assume the existence of certain
Bogomolnyi states and postpone most of our discussion of their origin until the
following sections. The bosonic massless fields of an $N\ge4$ supergravity
theory are the four-dimensional space-time metric $g_{\mu\nu}$, scalars $\phi
^i$ taking values in a sigma-model target space $M$ with metric $g_{ij}$ and
vector fields $A_\mu^I$ with field strengths $F^I_{\mu \nu}$. The gauge group
has  rank $k$ and is abelian for generic points in the moduli-space, in which
case $I=1,\dots,k$. The lagrangian, omitting fermions, is
$$
L=\sqrt
{-g} \left({1\over4}  R -\half g_{ij}(\phi)\partial_\mu\phi^i\partial^\mu\phi^j
-{1\over4}m_{IJ}(\phi) F^{I \mu\nu}F_{\mu\nu}^J - {1\over8}
\varepsilon^{\mu\nu\rho\sigma}a_{IJ}(\phi)F_{\mu\nu}^I F_{\rho\sigma}^J\right)
\eqn\one
$$
for some matrix functions $m,a$, with $m$ being positive definite. These
massless fields may couple to massive fields, which can be viewed as sources
for $(g,A,\phi)$. It is a feature of such theories that the mass of any field
configuration satisfies a classical bound of the form [\GH,\HT,\HL,\GKLTT]
$$
M^2 \ge \ZZ ^A \RR _{AB} (\bar \phi) \ZZ^B
\eqn\two
$$
where
$$
\ZZ = \pmatrix{p^I\cr q_I}
\eqn\three
$$
and $p$ and $q$ are the magnetic and Noether electric charges defined by
integrals over the two-sphere at spatial infinity, as in [\HT]. The matrix
$\RR$
is a function of the asymptotic values $\bar \phi^i$ of the scalar fields. In
the quantum theory, a similar bound applies to all quantum states, with the
numbers $\bar \phi^i$ now to be interpreted as the expectation values of the
scalar fields $\phi^i$, parameterising the possible vacua.

The charge vector $\ZZ$ satisfies the  DSZ quantization condition, which
implies that $q$ takes values in some lattice $\Gamma$ and $p$ takes values in
the dual lattice $\tilde \Gamma$. The matrix $\RR$ is a continuous function of
$\bar \phi$, so that the masses of Bogomolnyi states are also continuous
functions of $\bar\phi$. Under certain circumstances, to be elaborated shortly,
the matrix $\RR_{AB} (\bar \phi)$ has a (fixed) number of zero eigenvalues, for
all values of $\bar\phi$. This might make it appear that there should be extra
massless particles for all values of the moduli, but this is not the case for
two reasons. First, at any given point on moduli space there may be no points
in
the charge lattice that lie in the Kernel of $\RR_{AB} (\bar \phi)$. Second, as
we shall see in more detail later, not all points in the lattice of charges
allowed by the DSZ quantization condition actually occur in a given theory. In
particular, in string theory only those points in the electric
charge lattice that are consistent with the physical state conditions of
perturbative string theory can correspond to states in the string spectrum, and
there are no such points whose charges are in the kernel of $\RR _{AB}$ for
generic
points in moduli space. For special values of $\bar \phi$, however, a finite
number of string states  have charge vectors in the kernel, so that they become
massless. Conversely, the mass of a Bogomolnyi state with given charge vector
can vanish only for certain values of $\bar\phi$.  We shall return shortly to
consider the circumstances under which all conditions for massive Bogomolnyi
states to become massless at special points in moduli space can be satisfied in
a string theory, but we shall first examine some general consequences of $N=4$
or $N=8$ supersymmetry in the event that this phenomenon occurs.

We shall consider here only those Bogomolnyi states which preserve half the
supersymmetry of the vacuum. States which preserve some, but less than half,
the supersymmetry are also sometimes called `Bogomolnyi states' but they
saturate a stronger bound and will not be relevant to the phenomena to be
discussed here, for reasons to be explained below. Bogomolnyi states that
preserve half the supersymmetry fit into ultra-short $N=4$ or $N=8$ massive
supermultiplets with highest spin $h$; these have the same spectrum of helicity
states as the corresponding massless supermultiplets with highest spin $h$
(apart from the obvious charge doubling).
In a given vacuum,  the   ultra-short supermultiplets are precisely those that
saturate the bound \two.  However, as the moduli are varied, it is possible
that  states that fit into  $16$  ultra-short multiplets saturating \two\ for a
given point in    moduli space may combine into a long multiplet that does not
saturate \two\
for other points.  The number of ultra-short Bogomolnyi supermultiplets is
therefore conserved modulo $16$  (but note   that not every combination of
$16$
ultra-short supermultiplets can be combined into an unshortened supermultiplet
because the latter has highest spin $h\ge 2$ whereas an ultra-short
supermultiplet has highest spin $h\ge1$).

Thus, as $\bar \phi$ is continued to $\bar \phi_0$, the ultrashort Bogomolnyi
supermultiplets with charge vector ${\ZZ}_0$ must continue  (at least modulo
$16$)
to massless supermultiplets with the same highest spin. We do not expect the
new
massless supermultiplets to have highest spin $h\ge 2$, as these would lead to
well-known inconsistencies\foot{These inconsistencies might be avoided  if an
infinite number of supermultiplets become massless. This occurs in a
`decompactification limit' in which some compact dimensions become non-compact,
or in a null string limit. Such phenomena, which we will not consider here, are
associated with points on the boundaries of moduli space.}. Since all $N=8$
supermultiplets have highest spin of at least two we should not expect any
massive supermultiplets to become massless at special points in the moduli
space of compactifications that preserve $N=8$ supersymmetry, e.g. the
$T^6$ compactification of the type II superstring. We shall verify this
prediction below. For $N=4$ there remain  two possibilities: $h=1$ and
$h=3/2$.

Consider first the $h=3/2$ case\foot{This has been recently considered in
[\Kallosh].}. The existence of additional massless spin-$3/2$ states  implies
an enhanced $N>4$ local supersymmetry, but this is possible only if {\it all}
massless states belong to the graviton supermultiplet, since there are no
massless matter supermultiplets (with $h \le1$) for $N>4$. Moreover, the total
number of massless vectors would increase since the $N=4$ supermultiplet with
$h=3/2$ contains vector fields. In the cases of most interest  to us here, the
toroidally compactified heterotic string or type II on $K_3\times T^2$, the
number of massless vector fields at a generic point in the moduli space is
already 28, so that we would need an effective $N>4$ supergravity with more
than 28 vector fields. There is no such theory (the $N=8$ theory has exactly
28). Moreover, the gauge group of the massless vector fields would have to be
non-abelian, for reasons explained below, and it is difficult to reconcile this
with a vanishing cosmological constant in a pure supergravity theory. For these
reasons, we exclude the possibility of additional $h=3/2$ massless
supermultiplets. Since partially shortened supermultiplets saturating a
stronger bound must have highest spin $h\ge 3/2$, this exclusion explains why
we may restrict our attention to ultrashort multiplets.

This leaves the possibility that massive $N=4$ vector multiplets become
massless at special points in the moduli space of a compactification preserving
$N=4$ supersymmetry. Ultra-short massive vector multiplets come in central
charge doublets which couple to the vector field $A^0$ of the corresponding
central charge. Consider the case in which only one such charge doublet with
vector fields $A^+,A^-$ (and their superpartners) becomes massless. Since the
effective massless theory now contains three vector fields with a trilinear
$A^0A^+A^-$ coupling, consistency implies that the original $U(1)^k$ gauge
symmetry is enhanced to the non-abelian group $U(1)^{k-1}\times SU(2)$. Note
that there are also additional massless scalars, but that these have quartic
interactions (as required by $N=4$ supersymmetry) so that their expectation
values do not constitute new moduli. More generally, several charged doublets
may become massless simultaneously, leading to an enhanced symmetry
group of higher dimension. The rank, however, must remain equal to $k$ (and the
maximal rank of the maximal simple subgroup equal to $k-6$). This is because
each of the additional massless vector multiplets is charged with respect to
one of the $k$ original $U(1)$'s.

We now investigate the conditions under which the matrix $\RR$ has zero
eigenvalues. Consider first the case of $N=8$ supergravity, which can be
interpreted as the effective theory for the toroidally compactified type II
superstring since the moduli space of this compactification is
$E_7/(SU(8)/\Z_2)$ and it preserves $N=8$ supersymmetry. The scalar fields of
$N=8$ supergravity take values in the coset space $E_7/(SU(8)/\Z_2)$ and can be
represented by a $56\times 56$ matrix function $\V(x)$ taking values in $E_7$.
This transforms under rigid $E_7$ transformations with parameter $\Lambda$ and
local $SU(8)$ transformations with parameter $h(x)$ as
$$
\V (x)\rightarrow h(x)\V (x)\Lambda^{-1} \ .
\eqn\four
$$
The $56\times 56$ $\RR$-matrix for this theory is given by
$$
 \RR  = \V ^t \V\ ,
\eqn\five
$$
where $\V^t$ is the transpose of $\V$. For every point in the coset space
$E_7/[(SU(8)/{\Z}_2)]$, the matrix $\V$ is non-degenerate and so $\RR$ must
have non-zero determinant. This means that, as predicted, there can be no
points
at which $\RR$ has zero eigenvalues.

We now turn to $N=4$ supergravity coupled to $m$ vector multiplets. The number
of vector fields is $k=6+m$, since the supergravity multiplet contains six
vector fields. The scalars take values in the coset space $G/H$ where
$G=SL(2;\R)\times O(6,m)$ and $H=U(1)\times O(6)\times O(m)$. The scalars can
be
represented by a $2k\times 2k$ matrix function $\V (x)$ taking values in $G$.
In formulating the theory [\dR,\berg,\HT], it is also useful to introduce a
scalar-dependent $6\times k$ matrix $t$, of rank 6, which converts $SO(6,m)$
indices into $SO(6)$ indices. Introducing the $12 \times 2k$ matrix $K= \II
\otimes t$ where $\II$ is the $2\times 2$ identity matrix, the $\RR$-matrix
takes the form
$$
\RR = \V ^t K^t K \V\ .
\eqn\six
$$
As in the $N=8$ case, the determinant of $\V$ is non-zero but, by construction,
the rank of $K^tK$ is precisely 12, so $\RR$ has precisely $2m$ eigenvectors
with zero eigenvalue (when $m=0$, $K^tK$ is the identity matrix).

To investigate the nature of these zero-eigenvalue eigenvectors it is
convenient to split the modulus $\bar \phi = (\varphi, \lambda) \in G/H$ into a
complex coordinate $\lambda \in SL(2,\R)/U(1)$ and coordinates $\varphi \in
O(6,m)/O(6)\times O(m)$ and rewrite the $N=4$ Bogomolnyi mass formula in the
form [\SS,\Sen]
$$
M^2 = {1 \over 16}
\pmatrix {p & q}
[{\cal S} \otimes (M+L)]
\pmatrix{p \cr q}
\eqn\rhet
$$
where
$$
{\cal S} = { 1 \over \lambda _2}\pmatrix { | \lambda|^2 & \lambda _1 \cr
\lambda _1 & 1}
\eqn\sis
$$
is an $SL(2,\R)$ matrix depending on $\lambda = a_4+ie^{\Phi_4} =\lambda _1 + i
\lambda_2$, where $a_4$ and $\Phi_4$ are scalars that we shall refer to as the
four-dimensional axion and dilaton respectively. For the toroidally
compactified heterotic string, $\Phi_4$ comes from the ten-dimensional dilaton
$\Phi_{10}$, while for the type II string compactified on $K_3\times T^2$,
$\Phi_4$ comes not from the ten-dimensional dilaton but from one of the $T^2$
moduli. Here, $L$ is the invariant metric of $O(6,m)$ (with six eigenvalues of
$+1$ and $m$ of $-1$) and $M(\varphi)$ is an $O(6,m)$ matrix  whose explicit
form is given, for $m=22$, in  [\SS,\Sen].

As ${\cal S}$ is unimodular, any zero mass states in the spectrum correspond to
the zero eigenvalues of the $(6+m)\times (6+m)$ matrix $M(\varphi) +L$.
There is a special point in $\varphi$-space at which $M=\II$ (the identity
matrix). A general point $\varphi$ in moduli space is obtained by acting with
an
$O(6,m)$ transformation represented by a matrix $\Omega(\varphi)$, in which
case
$M(\varphi)=\Omega(\varphi)^t \Omega(\varphi)$. Thus $M+L=\Omega ^t(\II +L)
\Omega $ has $m$ zero eigenvalues and so the matrix ${\cal R}={\cal S}\otimes
(M+L)$ has 2m zero eigenvalues, for all values of $\bar\phi=(\lambda,\varphi)$,
in agreement withour analysis above. Furthermore, a given $(6+m)$-vector $V$
will be in the kernel of $M(\varphi) +L$ if and only if $\Omega(\varphi)V$ is
in
the kernel of $\II+L$. In particular, this requires the $O(6,m)$ norm,
$V^2\equiv V^IL_{IJ}V^J$, of $V$ to be strictly negative. Thus, an electrically
charged Bogomolnyi state with $q=V$ and $p=0$ can never become massless if
$V^2\ge0$, whereas if $V^2<0$ it will be massless for precisely those values of
$\varphi$ for which $\hat V=\Omega(\varphi)V$ is in the kernel of $\II+L$.
Moreover, for any $V$ with $V^2<0$, there will be  values of $\varphi$ for
which
$V$ represents a massless state, since there will be $O(6,m)$ transformations
that rotate $V$ into the kernel of $\II+L$.

Instead of dealing with electric charge vectors  $q \in \Gamma$, it is
sometimes useful to consider the charge vectors $\hat q \in \hat \Gamma$ where
$\hat q = \Omega(\varphi)q$ and $\hat \Gamma $ is the lattice obtained from
$\Gamma $ by acting  with the $O(6,m)$ transformation $\Omega(\varphi)$. For
electrically charged states, the mass formula \rhet\ is
$$
M^2 = {1 \over 16 \lambda _2} \hat q ^t (\II+L)\hat q
\eqn\rheto
$$
so that instead of considering fixed charges and varying the matrix $M$, we can
vary the lattice $\hat \Gamma$ and keep the lattice metric fixed. We are
interested in those points in the lattice $\hat \Gamma$  that also lie in the
kernel of $\II +L$. It can happen that $\hat \Gamma$ has no basis vectors that
are annihilated by $\II +L$, in which case no massless states arise for that
value of $\varphi$. If $\hat \Gamma \cap {\rm ker}(\II +L)$ is non-trivial, the
next question to be addressed is whether the potential massless states arise in
a given theory, since not all points in $\hat \Gamma$ need correspond to
physical states. The lattice $\hat \Gamma$ of allowed charges can be decomposed
into $O(6,m;\Z)$ orbits, each of which has a particular
value of the $O(6,m)$ norm $q^2= \hat q^2$. As we have just seen, only those
orbits with $q^2<0$ can give rise to massless states. In string theory, most
states have $q^2 \ge 0$  so that only a  small part of the spectrum can  give
rise to extra massless states in this way.

 From the mass formula \rhet, it is clear that if an electric Bogomolnyi state
with charge $(p,q) =(0,V)$ becomes massless at some special modulus
$\varphi_0$, then any state with charge vector $(p,q)=(mV,nV)$ with $m,n$
integers will have a mass that also tends to zero as $\varphi \to \varphi_0$.
This fact has some interesting implications for string theory
compactifications,
which we now examine in further detail.


\chapter{Application to Superstring Compactifications}

We shall now show how the results just obtained apply to specific superstring
compactifications. For both the generic $T^6$ compactification of the heterotic
string and the $K_3\times T^2$ compactified type II superstrings the effective
four-dimensional field theory is N=4 supergravity coupled to 22 abelian vector
supermultiplets, so the discussion of section 2 applies with $m=22$. In each
case the scalar field sigma-model of the effective four-dimensional field
theory
has a target space given by the product of $O(22,6)/[O(22)\times O(6)]$,
parameterised by $\varphi$, and $SL(2,\R)/U(1)$, parameterisd by $\lambda$. The
equations of motion of the supergravity theory are invariant under the
continuous symmetry group
$$
SL(2;\R)\times SO(6,22)
\eqn
\dual
$$
Quantum effects break this to the discrete group
$$
SL(2;\Z)\times SO(6,22;\Z)\ .
\eqn
\STdual
$$
For both the $T^6$ compactified heterotic string and the $K_3\times T^2$
compactified type II superstring this discrete group is conjectured to be a
symmetry of the spectrum. In the case of the heterotic string, $SO(6,22;\Z)$ is
the T-duality group and $SL(2;\Z)$ is the S-duality group. As we shall see
later, for the type II string the situation is less straightforward:
the T-duality group of perturbative symmetries contains $SL(2;\Z)$    and an
$SL(2;\Z)\times O(4,20;\Z)$ subgroup of $O(6,22;\Z)$, while the $SL(2;\Z)$
S-duality group acting on the dilaton is also a subgroup of $O(6,22;\Z)$.
The full symmetry \STdual\ was conjectured in [\HT] to be a symmetry of the
non-perturbative spectrum and  has recently been investigated further in
[\AM]. Note that although T-duality is known to be a symmetry of the
perturbative spectrum its status as a symmetry group of the full
non-perturbative spectrum remains conjectural.

Consider the $T^6$ compactified heterotic string. In this case the parameters
$\varphi$ are the $T^6$ moduli and the real and imaginary parts of the complex
variable $\lambda$ are the constant values of the axion and dilaton fields
[\SS]. For weak coupling, i.e. $g_4 <<1$, where $g_4 \equiv \left\langle
e^{\Phi
_4} \right\rangle$ and is the same as the ten-dimensional string coupling
$\left\langle e^{\Phi _{10}} \right\rangle $ for the heterotic string on $T^6$,
the electrically charged Bogomolnyi states arise as modes of the fundamental
string and have charges $q$ for which $q^2$ is even and satisfies
$q^2 \ge -2$. We have shown that states with $q^2 \ge 0$ do not become
massless, which is just as well since that would have meant higher-spin
supermultiplets becoming massless. Also, the $q^2=-2$ states indeed fit into
vector supermultiplets, in agreement with our earlier analysis. Thus, only
Bogomolnyi states with $q^2=-2$ can become massless and for a given value of
$\varphi$, the extra massless states are the ones satisfying $q^2=-2$ and such
that $\hat q \in {\rm ker}(\II +L)$. For generic values of $\varphi$, there
will
be no such massless states, but for special values there will a finite number
of
vectors in the charge lattice satisfying these conditions, and these can be
identified with the root vectors of the enhanced gauge algebra [\Narain].
 Conversely, the general analysis of section 2 shows that for a given charge
vector satisfying these conditions there is always a vacuum for which this
happens.

Given any Bogomolnyi vector supermultiplet with $(p,q) =(0,V)$ such that
$V^2=-2$, there are also supermultiplets with $(p,q) =(V,nV)$ and $(p,q)
=(2V,nV)$ which are represented for weak coupling and at generic points in the
moduli space by BPS monopoles and BPS dyons [\Sen,\ssen,\HL]. It has been shown
[\SS,\Sen] that the conjectured $SL(2,\Z)$ symmetry of the heterotic string
spectrum implies Bogomolnyi states with $(p,q)=(mV,nV)$ for all co-prime
integers $m,n$. We saw in section 2 that all these states must become massless
together. Thus, S-duality and N=4 supersymmetry imply that as an enhanced
symmetry point of the heterotic string is approached there is an infinite set
of dyon states whose masses tend to zero, in addition to the purely electric
and
purely magnetic states. The interpretation of the magnetic and dyon states as
due to quantization of solitons presumably fails at points of enhanced symmetry
because the sizes of the monopole and dyons approach infinity as their mass
approaches zero\foot{It is possible that they may be considered as quanta of
wave solutions obtained by boosting a soliton solution to the speed of light
while keeping the mass fixed.}. This phenomenon is presumably another
indication of a phase transition to a special type of non-abelian Coulomb
phase.

It is instructive to explore further the consistency of the conjectured
S-duality [\FILQ,\SS,\Sen] of the four-dimensional heterotic string with the
phenomenon of symmetry enhancement. At strong coupling, $g_4 >>1$, the
four-dimensional heterotic string is conjectured to be related to the weakly
coupled theory by S-duality, with the roles of electric and magnetic charges
interchanged. If so, then at generic points in moduli space, electric states
with $(p,q) =(0,V)$ are perturbative string states for weak coupling and
represented at strong coupling by solitons of the dual theory, while magnetic
states with $(p,q) =(V,0)$ are solitons of the weakly coupled theory but are
perturbative states of the dual theory. It is expected that the theory can be
smoothly continued in $g_4$ without encountering any phase transition,
in which case S-duality implies that the electric and magnetic charges should
be on exactly the same footing. This would mean, in particular, that
magnetically charged vector states become massless at strong coupling through
the HFK mechanism in the dual theory. S-duality implies that magnetically
charged states occur as perturbative states of the dual strong-coupling theory,
and if, as we are assuming, these can be continued in $g_4$ back to
magnetically
charged states at weak coupling, it is clear that magnetic states with masses
tending to zero at the special points must be present in the weakly-coupled
theory, even though their perturbative description as solitons in the weak
coupling theory breaks down.

The string theory coupling constant is $g_{10}=\left\langle e^{\Phi_{10}}
\right\rangle $, but the dilaton $\Phi_{10}$ appears differently in the two
theories, as we shall now argue. In particular, $\Phi_{10}$ occurs in the
$SL(2;\R)/U(1) $ coset space for the heterotic string (and so can identified
with $\Phi_4$) while the type II dilaton lies in the $ O(6,22)/O(6)\times
O(22)$ coset space after compactification on $K_3 \times T^2$. This means that
perturbative effects in the heterotic string can be non-pertiurbative in the
type II string, and vice versa. To see this, we shall  focus on the subgroup
$O(4,20)\times SL(2;\R)_S \times SL(2;\R)_T \times SL(2;\R)_U$ of \dual. For
both superstring compactifications, $SO(2,2) \sim SL(2;\R)_T \times SL(2;\R)_U$
acts on the moduli space of $T^2$ and $SL(2;\R)_S$ acts on the dilaton and
axion fields arising in the usual way from the $D=10$ dilaton $\Phi_{10}$ and
the antisymmetric tensor gauge field. The space $O(4,20)/[O(4)\times O(20)]$,
modulo the discrete group \STdual, is the moduli space for $K_3$ and for the
Narain construction of heterotic strings in six dimensions. For the heterotic
string, $SL(2,\R)_T \times SL(2,\R)_U \subset O(6,22)$. For the type II string,
however, the RR charges are inert under S-duality [\HT] and do not couple to
the
ten-dimensional dilaton $\Phi_{10}$ [\bergort,\Witten] (see also [\ferras]), so
that $\Phi_{10}$ cannot be identified with $\Phi_{4}$, which does couple to all
charges. Thus it must be the  case that $SL(2,\R)_S \subset O(6,22)$, and in
fact $SL(2,\R)_S \times SL(2,\R)_U \subset O(6,22)$, so that the $SL(2,\R)_S$
and $SL(2,\R)_T$ factors are interchanged compared with the heterotic string,
as in [\duff]. In particular, for the heterotic string, $\Phi_4$
coincides with the $D=10$ dilaton $\Phi_{10}$ while for the type II string they
are distinct: $\Phi _{10}$ lies in $O(6,22)/[O(6)\times O(22)]$ while $\Phi_4$
is one of the $T^2$ moduli. It is interesting to note that the equivalence of
the heterotic and type II superstring compactifications together with T-duality
in each theory implies invariance of the spectrum under the full group
$O(4,20)\times SL(2;\R)_S \times SL(2;\R)_T \times SL(2;\R)_U$ since
 what is non-perturbative in one is perturbative in the other.

The main implication of section two for the $K_3\times T^2$ compactification of
the type II superstring is that symmetry enhancement will occur if the
Bogomolnyi spectrum includes electrically charged states with $q^2 <0$. These
states do not occur in perturbation theory but they may appear in the
non-perturbative spectrum as Ramond-Ramond (RR) solitons [\HT]; we shall
address
this question in the following section. Note, however, that if the conjectured
equivalence to the heterotic string is correct the type II string studied using
perturbation theory in the $T^2$ modulus
$g_4\equiv \left\langle e^{\Phi_4} \right\rangle$ should give the same results
as the heterotic string expanded in the usual way order by order in the string
coupling $g$. For the type II string on $K_3\times T^2$, the usual string
perturbation theory in $g$ is not useful as the RR soliton effects are
non-perturbative in $g$, whereas an expansion in $g_4$ should yield results
equivalent to those of the usual perturbative heterotic string.


\chapter{Bogomolnyi states}

We have seen in the preceeding sections that symmetry enhancement in
superstring theories need not be associated with toroidal compactifications.
The
fact that such an association has been made in the past is simply a reflection
of the fact that this is the only example that can be discovered by
conventional
perturbative string theory methods. Fortunately, N=4 supersymmetry is such a
strong requirement that in this case it is not necessary to solve the strongly
coupled theory to obtain the information needed to determine whether a given
N=4
supersymmetric theory will exhibit non-perturbative symmetry enhancement.
If the existence of a Bogomolnyi state with a given charge vector can be
established for some value of the coupling constants, then there will be a
state
with that charge vector for all values of the coupling constants, with mass
given by the Bogomolnyi mass formula. In this section we will discuss the
Bogomolnyi spectrum.

In the context of the purely massless effective four-dimensional $N\ge4$
supergravity theory the  Bogomolnyi states breaking half the supersymmetry can
only arise from quantization of extreme black holes [\GETC,\Ort,\Kalort,\GP]
for
special values of the `dilaton coupling constant', $a$. Specifically, they
arise
from $a=\sqrt{3}$ extreme black holes of the $N=8$ supergravity [\HT] and from
$a=\sqrt{3}$ and $a=1$ extreme black holes of $N=4$ theories [\Druff ]. One
obstacle to the interpretation of quantized extreme black holes as Bogomolnyi
states is the fact that only the $a=1$ magnetic ones are completely
non-singular. The way around this obstacle in the context of string theory is
that the four-dimensional supergravity is interpreted as the effective theory
of a compactified D=10 theory. In this case one can interpret the extreme black
hole solutions of the four-dimensional theory as either (i) Kaluza-Klein modes
of the massless D=10 fields, (ii) their magnetic duals, the KK monopoles
[\SGP], or (iii) as `wrapping modes' of the   $p$-brane solitons [\HT]. We
shall
argue that all the Bogomolnyi states needed for symmetry enhancement arise from
non-singular, or otherwise physically acceptable, solutions of the D=10 theory
describing weak coupling or of the dual effective theory describing strong
coupling.

Let us first consider how higher dimensions help for the heterotic string. The
$a=\sqrt{3}$ magnetic extreme black holes can now be interpreted either as
Kaluza-Klein monopoles (for those whose magnetic field is of KK origin) or as
compactified five-branes of the 10-dimensional theory (for those whose magnetic
field is of antisymmetric tensor origin). Some of the latter are interpreted as
BPS monopoles in four dimensions [\HL,\GHL]. Thus, all of the magnetic extreme
black holes breaking half the $N=4$ supersymmetry are non-singular in the
ten-dimensional context and it should therefore be possible to determine the
corresponding quantum states by semi-classical methods at weak string coupling.
The resulting non-perturbative magnetic states are therefore in the string
spectrum for weak string coupling and because they belong to ultra-short
supermultiplets they must remain in the spectrum (at least modulo  $16$) at
strong
coupling. The singularities of the electric extreme black holes cannot be
completely removed in this way, but this difficulty can be resolved by
identifying them with the perturbative Bogomolnyi states of the heterotic
string. The particles in the perturbative string spectrum have electric and
gravitational fields and might therefore be expected to appear as extreme black
holes in the effective field theory. Indeed, the electric extreme black holes
carry the same quantum numbers as the Bogomolnyi states in the perturbative
heterotic string spectrum. Finally, such an identification is expected from
considerations of S-duality [\Druff,\HT], and corresponds to identifying the
solitonic string or one-brane solution in ten dimensions with the fundamental
string [\HT].

According to this picture, extreme black holes that break half the
supersymmetry correspond either to fundamental string states or are the
projection to four dimensions of KK monopoles or $p$-brane solitons that are
non-singular in a higher dimension. This was implicit in our previous work
[\HT]
on U-duality in type II string theory, but since there we passed over issues
related to the singularities of $p$-brane solitons of type II superstrings we
now take another look.

The electrically and magnetically charged extreme black holes of the type II
string compactified on either $T^6$ or $K_3 \times T^2$ can be interpreted as
KK
monopoles, or as compactified $p$-branes of either the type IIA or the type IIB
ten-dimensional supergravity [\HT]. Recall that these consist of an electric
one-brane and magnetic five-brane in the Neveu-Schwarz/Neveu-Schwarz (NS-NS)
sector and, in the Ramond-Ramond (RR) sector, $p$-branes with $p=0,2,4,6$ for
type IIA and $p=1,3,5$ for type IIB. Of these, only the NS-NS fivebrane and the
type IIB self-dual threebrane is completely non-singular in string sigma-model
variables. The NS-NS one-brane is to be identified with the fundamental string,
as for the heterotic string. The type IIA magnetic $p$-brane solitons are all
completely non-singular as solutions of 11-dimensional supergravity
[\GHT,\PKT], which can be interpreted as the effective action of the type IIA
superstring at strong coupling [\Witten]. This justifies consideration of all
magnetic extreme black holes in the effective four-dimensional theory, but this
is not sufficient by itself because the non-perturbative string states expected
on the basis of extreme black hole solutions include some {\sl electrically}
charged ones from the Ramond-Ramond (RR) sector [\HT]. The inclusion of these
states (required by U-duality [\HT]) can also be justified by considerations of
eleven dimensional supergravity to the extent that the 0-brane RR solitons can
be identified with KK modes of D=11 supergravity compactified on a circle
[\PKT,\Witten], and the RR 2-brane soliton has its origin in the 11-dimensional
membrane soliton. In fact, apart from the KK modes and the KK monopoles, all
$a=\sqrt{3}$ extreme black holes solutions of $N=8$ supergravity in four
dimensions can be regarded [\HT] as `compactifications' of either the 2-brane
[\DS] or the 5-brane [\Gu] soliton of 11-dimensional supergravity,
both of which are needed for U-duality of the $T^6$-compactified type II
superstrings. The problem that remains to be confronted is that whereas
the 11-dimensional 5-brane soliton is completely non-singular, the 2-brane
soliton  has a time-like singularity hidden by an event horizon [\DGT] (as for
the four-dimensional extreme Reissner-Nordstrom solution). This suggests {\it
either} that we consider this type of singularity to be consistent with the
apellation `soliton', {\it or} that we identify the 11-dimensional 2-brane
soliton with a fundamental 11-dimensional supermembrane.

We now turn to the type IIB superstring.
The NS-NS sector string and five-brane are the same as those already discussed
for the type IIA string while   the RR sector solitons that break half the
supersymmetry consist of a
string, a five-brane, and a
self-dual three-brane.
 The self-dual three-brane solution is geodesically
complete [\GHT], but the RR string and five-brane solitons are not (the
solutions
may be found in [\PKT,\Oopen]).  The weakly coupled theory justifies inclusion
of the charges corresponding to the NS-NS string (which is identified with the
fundamental
string) and the NS-NS five-brane  and RR three-brane (which are non-singular
solitons).
As for the type IIA string, the inclusion of the other charges is justified by
the fact that they are carried by acceptable solutions of the low-energy
effective  theory at strong  coupling.
In this case, the conjectured strong coupling limit is a dual type IIB string
theory [\Witten] with NS-NS and RR charges interchanged. As will be argued
below, the RR  five-brane is a non-singular soliton of the dual theory while
the RR string should be identified with the dual fundamental string.
Thus the inclusion of all types of charge  is justified in the type IIB theory.

It was conjectured in  [\HT] that this theory
has an  $SL(2;\Z)$ duality symmetry. This includes a $\Z _2$ subgroup which
 interchanges weak and strong coupling regimes and
it was this that led to the conjecture  that the type IIB theory, like the
four-dimensional heterotic string, is
self-dual. This $\Z_2$ duality
corresponds to  a symmetry of the equations of motion of the IIB supergravity
theory that
interchanges the NS-NS with the RR solutions, so the RR string and fivebrane
of the weakly coupled theory become the NS-NS string and fivebrane of the
effective strongly coupled theory, as required.

Semi-classical quantization of the solitons in D=4 that can be interpreted as
wrapping modes of the $p$-brane solitons in D=10 or D=11 will lead to
supermultiplets of massive states. Since these $p$-brane solitons preserve half
the supersymmetry, the ground state soliton supermultiplet will preseve
half the supersymmetry of the vacuum state, which implies that these
supermultiplets are ultra-short ones, as required for our arguments of section
2. This remains true as the coupling constants and the moduli of the
compactification are varied, so that the multiplets associated with   the
$p$-brane solitons remain ultra-short and do not combine into longer
multiplets. This in turn implies that the resulting four-dimensional states
indeed saturate the bound \two. Certain of these will then become
massless at the special points in moduli space, as we shall argue below.

What the above arguments show is that each abelian gauge field in the effective
four-dimensional $N=4$ supergravity couples to a tower of massive charged
states, irrespective of which superstring theory this is the effective action.
This result can be considered to be a generalization, first, of KK theory (for
which one finds only the KK states) and, second, of perturbative string theory
(for which one finds only the KK states and the winding states). According to
either KK theory or perturbative string theory, certain gauge fields (those
coming from the metric or the two-form potential) are singled out for special
treatment. From the new perspective, in which one also includes
non-perturbative $p$-brane `wrapping' modes, {\it all} abelian gauge fields are
on the same footing. This is in accord with the conjectured S-duality [\SS] and
U-duality [\HT] of the heterotic and type II strings.

Having established the existence of certain Bogomolnyi states, we now turn to
the question of which of these can become massless and so lead to symmetry
enhancement. Consider first the compactification of the type IIA string on
$K_3$. The resulting six dimensional string theory will have Bogomolnyi states,
identifiable in the semi-classical regime as wrapping modes of the type IIA
2-brane around the homology 2-cycles of $K_3$ [\HT]. There are special points
in the moduli space of $K_3$ at which the $K_3$ surface becomes singular due
to the collapse of certain homology 2-cycles: the area of the 2-cycles tends
to zero as
the special point is approached. The mass of a state obtained by wrapping
a 2-brane round one of these cycles is proportional to the area of the cycle,
so that it is a natural candidate for a state that becomes massless at a
special point in the full moduli space of metrics and antisymmetric tensors on
$K_3$\foot{As pointed out by Aspinwall (hep-th/9507012), the extra
antisymmetric tensor moduli must be adjusted in order to ensure that a point at
which a two-cycle collapses corresponds to a point of enhanced symmetry. In the
following, we shall assume that this adjustment has been made.}.
In [\Witten], it was shown that there is a one-to-one correspondence between
the points in the Narain moduli space of toroidally compactified  heterotic
strings in six dimensions at which a non-abelian symmetry group $G$ emerges
and the points in the $K_3$ moduli space at which there is a singularity of
type $G$. At such a singularity, $r$ two-spheres collapse to a point, where
$r$ is the rank of the group. Each two-sphere is associated with a simple
positive root of $G$, so that exactly the right number of states are becoming
massless. The vector fields taking values in the Cartan sub-algebra of $G$ are
perturbative states of the theory, so the situation is reminiscent of the HFK
mechanism: there, the Cartan sub-algebra is associated with perturbative
states, while the extra generators of  enhanced gauge symmetry are
non-perturbative. The arguments of [\Witten] then imply that the Narain lattice
of the heterotic string can be identified with the lattice of integral points
in the real cohomology of $K_3$. Furthermore, the intersection form is
given by the the Dynkin diagram of $G$, and this seems to be the source
of the non-abelian interactions.

The states becoming massless can be reliably interpreted as wrapping modes for
large mass when the semi-classical approximation is valid, so that in this
regime the area of the two-cycle is given by the Bogomolnyi mass formula.
As the mass tends to zero, it is not clear whether or not the semi-classical
picture of the state as a wrapped brane remains tenable. Nonetheless, the
Bogomolnyi mass formula remains reliable, and can be expected to still give
the area of the cycle. Thus the state will become massless as the
two-cycle collapses, irrespective of whether the semi-classical approximation
is valid near the point at which the cycle collapses. At these special points
in moduli space, there is symmetry enhancement of the six dimensional
string theory, with the product of $U(1)^4$ and a rank 20 non-abelian group as
the gauge group. This feature can be preserved by a subsequent
compactification
to four dimensions on $T^2$, which will allow extension of the gauge group to
$U(1)^6$ times a rank 22 non-abelian group. Some of the extra symmetries come
from  the perturbative HFK mechanism. The enhanced symmetry groups  of the
toroidally compactified heterotic string and the type II string compactified on
$K_3\times T^2$ are the same at corresponding points of the two moduli spaces,
and this implies that at least those parts of the Bogomolnyi spectra  of the
two theories that become massless somewhere in moduli space must agree.

We can now provide an explanation, in the context of type II compactifications,
of why each electrically charged Bogomolnyi state that becomes massless in
special vacua is accompanied by a magnetically charged state. Consider again
the
compactification of the type IIA superstring theory to D=6; as explained in
[\HT], in addition to the 22 electrically charged states resulting from
wrapped two-branes, the type IIA four-brane yields 22 `elementary' two-branes
in
D=6 because of the 22 independent homology two-cycles of $K_3$ around
which it can wrap. On subsequent reduction on $T^2$ each of these D=6
two-branes can wrap around the two-torus to produce a magnetically charged
state
of the four dimensional theory. There is one such magnetic four-brane wrapping
mode for each electric membrane wrapping mode, as one would expect from the
fact
that the electric dual of a membrane in D=10 is a fourbrane. Moreover, it is
clear from the above description of the D=10 origin of these states that each
homology two-cycle of $K_3$ is associated with electric-magnetic pairs of
Bogomolnyi states which must become simultaneously massless as the area of the
two cycle vanishes. This result is in perfect accord with our earlier
conclusion
to this effect based solely on N=4 supersymmetry. Precisely for this reason,
the result is true also for the toroidally compactified heterotic string, as
mentioned earlier, although the mechanism for this remains obscure. Curiously,
although the electric part of the symmetry enhancement mechanism is better
understood for the heterotic string, it seems  easier to understand the
magnetic part of this mechanism for the $K_3\times T^2$ compactified type II
string. It would be interesting to understand the extra massless dyon states
from this point of view as well.


\chapter{Symmetry enhancement in D=11 supergravity compactifications}

It has been conjectured [\Witten] that the effective action for the strongly
coupled D=7 heterotic string, at generic points in its moduli space, is the D=7
field theory found by $K_3$ compactification of D=11 supergravity. Here we
shall provide further evidence for this conjecture by showing how the symmetry
enhancement known to occur for the heterotic string at special points of its
moduli space also occurs for D=11 supergravity.

The effective D=7 field theory obtained by $K_3$ compactification of D=11
supergravity is an N=2 (minimal) supergravity theory coupled to 19 vector
supermultiplets. Since there are three vector gauge potentials in the
supergravity multiplet there are 22 vectors in all, and the gauge group of the
effective D=7 field theory is $U(1)^{22}$ (it is an abelian group of rank 22
but for reasons explained in [\HT] the charges are quantized so the group is
compact). There is no need to go into the details of this compactification
because the D=7 Maxwell/Einstein supergravity is completely determined by
supersymmetry once one specifies the number of vector supermultiplets [\BKS]; a
summary of the features relevant to the current discussion can be found in
[\PKTb]. From the standpoint of Kaluza-Klein theory these gauge potentials have
no massive modes to which they might couple but they do couple to the wrapping
modes of the two-brane soliton of D=11 supergravity around the 22 homology
two-cycles of $K_3$. These massive modes are charged with respect to the gauge
fields which  are obtained from the three-form potential of D=11 supergravity.
They are massive  Bogomolnyi states carrying D=7 central charges, the  D=11
dimensional origin of which is the two-form charge appearing in the
D=11 superalgebra [\AGIT]. At points in the $K_3$ moduli space at which a
two-cycle collapses, the associated Bogomolnyi  state will become massless for
reasons explained in the previous section\foot{Note that in this case the full
moduli space is that of $K_3$ metrics, so that no adjustment of additional
variables is necessary}. Thus, the symmetry enhancement
mechanism for compactification of D=11 supergravity on $K_3$ to D=7 is an
intrinsically `membrany' mechanism, analogous to the intrinsically `stringy'
mechanism for $T^3$ compactification of the heterotic string. This is in
accord    with the D=7 string/membrane duality picture elaborated in [\PKTb].

The D=11 fivebrane soliton yields 22 three-brane solitons in D=7 after wrapping
around the homology two-cycles of $K_3$. These are the duals of the membrane
wrapping modes just discussed. On further reduction to four-dimensions on $T^3$
each of these produces 22 magnetically charged Bogomolnyi states, so that we
again find magnetic partners to each electrically charged state whose mass
vanishes as a homology two-cycle of $K_3$ vanishes. As discussed above, the
phenomenon of magnetic charges becoming massless simultaneously with electric
ones is presumably indicative of some special non-abelian Coulomb phase. One
might have supposed that such a phenomenon would be restricted to four
dimensions since it is presumably related to infrared divergences of unconfined
massless non-abelian gauge fields. However, we see from this example that
symmetry enhancement in higher dimensions is related to an even stranger
phenomenon. In D=7, for example, each electric state that becomes massless is
partnered by a three-brane whose tension, and mass per unit volume, goes to
zero, i.e. we have a {\it null} three-brane. This is an example of a phenomenon
that happens much more generally: in many string theories and supergravity
theories, there are   points in moduli space at which $p$-brane solitons become
null, generalising the phenomenon of black hole solitons becoming massless
discussed here. Indeed, the higher dimensional origin of black hole solitons
becoming massless is often in $p$-brane solitons that become null.

\vskip 0.5cm
\noindent{\bf Acknowledgements}: We are grateful to Sergio Ferrara for
explaining to
us the results of [\Cer] and for suggesting that they might be relevant to the
$K_3\times T^2$ compactified type II superstring considered in [\HT].
We would also like to thank Jerome Gauntlett, Michael Green, John Schwarz and
Nathan Seiberg  for helpful discussions.


\refout
\bye


\refout
\bye